**Preprint vs 1**

# Structural and optical properties of in situ Eu-doped ZnCdO/ZnMgO superlattices grown by plasma-assisted molecular beam epitaxy

Anastasiia Lysak, [a] Aleksandra Wierzbicka, [a] Sergio Magalhaes, [b] Piotr Dłużewski, [a] Rafał Jakieła [a], Michał Szot, [a,c], Zeinab Khosravizadeh, [a] Abinash Adhikari, [a] Adrian Kozanecki [a] and Ewa Przeździecka *[a]

In situ Eu-doped ZnCdO/ZnMgO superlattices (SLs) with varying ZnCdO:Eu and ZnMgO sublayers thicknesses were deposited by plasma assisted molecular beam epitaxy. The influence of Eu doping into ZnCdO quantum wells on structural and optical properties of superlattices were investigated using X-ray diffraction (XRD), transmission electron microscopy (TEM), secondary ion mass spectrometry (SIMS), cathodoluminesence (CL) with a scanning electron microscope (SEM), and UV–Vis spectroscopy. Cross-sectional high resolution TEM images and SIMS depth profile measurements confirmed the high quality of as-grown ZnCdO:Eu/ZnMgO periodic structures. XRD analysis revealed the single-phase ZnO wurtzite structure of the samples. The CL spectra of both as-grown and annealed {ZnCdO/ZnMgO}$_{22}$ SLs doped with Eu show the dominant near-band-edge peak as well as the broad band attributed to deep-level emission. The characteristic emission peak at ~616 nm, corresponding to the $^5D_0 \rightarrow {^7F_2}$ transitions of $Eu^{3+}$ ions was observed. Furthermore, the intensity of the red luminescence increased after annealing at 700°C, while a subsequent increase in the annealing temperature leads to a reduction in intensity. The analysis of CL spectra showed that native defects are involved in energy transfer from the ZnCdO host matrix to $Eu^{3+}$ ions.

## 1. Introduction

Wide-band gap (WBG) semiconductors doped with rare earth (RE) elements such as Ce, Er, Eu, Tb, etc., have become the subject of considerable attention due to their unique properties and potential for use in a wide range of applications[1–4]. Among RE elements, Europium (Eu) exhibits distinctive luminescence characteristics, in particular red emission at ~615 nm, associated with the intra 4f-shell transition between the $^5D_0$ and $^7F_2$ states[5]. However, the efficiency of Eu luminescence strongly depends on the host matrix[4,6]. One of the promising host materials for doping with Eu is zinc oxide (ZnO), due to its unique properties such as a large direct band gap of 3.37 eV and a high exciton binding energy of 60 meV[3,7]. The band gap energy of ZnO can be modulated from visible to deep UV region by creating ternary solid alloys (ZnCdO, ZnMgO, ZnBeO, etc.). In this way, it is possible to control the optical properties of ZnO-based materials[8,9]. While most researchers focus on investigation of Eu-doped ZnO[3,5,10–15], the properties of ZnO-based quantum structures doped with Eu ions are poorly studied[16,17]. Based on the literature, it is well known that Eu-doped AlGaN/GaN quantum structures demonstrated an increase in the red emission efficiency compared to III-V bulk materials[4,18–20] thus studying ZnO-based quantum structures doped with Eu may be more suitable for future optoelectronic applications.

Molecular beam epitaxy (MBE) is one of the suitable methods for deposition of ZnO-based quantum structures. This method allows precise control of the composition and sublayer thicknesses[21,22]. Additionally, it allows for the introduction of doping during the growth process into the entire structure or only into quantum wells or barriers[23]. It is known that ion implantation of quantum structures can cause deformation and/or diffusion intermixing of sublayer materials, thereby increasing the number of defects in the samples[24]. This can be avoided by using the in-situ doping technique.

In this work, the structural and optical properties of in-situ Eu-doped {ZnCdO/ZnMgO}$_{22}$ superlattices (SLs) grown by plasma assisted MBE were investigated. The influence of annealing at different temperatures on the optical properties of SLs has been studied.

## 2. Experimental

In-situ Eu-doped ZnCdO/ZnMgO superlattices (SLs) with varying sublayer thicknesses were grown on commercially available *m*-plane sapphire (10.0) substrates by plasma assisted molecular beam epitaxy (PA-MBE) in a Riber Compact 21 system. Before growth, the *m*-plane sapphire ($Al_2O_3$) substrates were chemically cleaned in a $H_2SO_4:H_2O_2$ (1:1) mixture. Thermal purification of the substrates was carried out at a temperature of 150°C for 1 hour in the loading chamber and then the substrates were transferred to the growth chamber where they were annealed in vacuum at 700°C and then for 30 minutes in an oxygen ($O_2$) plasma. During the growth process, the radio-frequency (RF) power of the oxygen plasma was fixed at 240 W with an $O_2$ gas flow rate of 3 sccm. All in-situ Eu-doped ZnCdO/ZnMgO SLs were grown at 360°C (temperature measured with a thermocouple). High purity Zn (6 N), Mg (6 N), Cd (6 N), and Eu (4 N) elements were used as sources from the Knudsen effusion cells. The fluxes of Mg, Cd and Eu were $5.9 \cdot 10^{-8}$, $9.0 \cdot 10^{-8}$ and $0.5 \cdot 10^{-8}$ Torr, respectively. Varying sublayer thicknesses in the SL structures were achieved by different temperatures of the Zn effusion cell. The zinc flux was $3.7 \cdot 10^{-7}$ and $2.5 \cdot 10^{-7}$ Torr for samples A and B, respectively. Eu doping was done during the growth process into ZnCdO quantum wells (QWs). The ZnCdO:Eu and ZnMgO sublayers thicknesses ($h$) were controlled by adjusting the deposition time (2 minutes each). The number of sublayer pairs in the periodical structures was 22. The resulting structures corresponding to sample A was {ZnCdO:Eu$_{13nm}$/ZnMgO$_{12.5nm}$}$_{22}$ and for sample B: {ZnCdO:Eu$_{10.5nm}$/ZnMgO$_{10nm}$}$_{22}$.

Post-growth thermal treatment of Eu-doped {ZnCdO/ZnMgO}$_{22}$ SLs was carried out in a rapid thermal processing (RTP) system (AccuThermo AW610 from Allwin21 Inc.) at different temperatures (700°C, 800°C and 900°C) in an oxygen atmosphere for 1 minute.

The structural properties of these periodical structures were investigated by high resolution X-ray diffraction (HR XRD) using an X'Pert Pro-MRD Panalytical diffractometer with CuKα1 radiation. The diffraction patterns were collected over 28°–118°. The 10.0 2θ-ω scans were simulated utilizing the MROX 2.0 software, a widely recognized tool for simulating these scans based on the dynamical theory of X-ray diffraction for various Bravais lattices such as cubic, hexagonal, tetragonal, orthorhombic, and monoclinic structures (lattices)[25,26]. The software incorporates instrumental broadening effects by employing a Pseudo-Voigt function (PV) and Kalfa2 residuals as a secondary PV in the diffractograms. The fitting process was achieved through combinations of Genetic and Pattern Search Algorithms[25,26]. The as-deposited SLs were studied by transmission electron microscopy with use of a Titan 80-300 Cube microscope operating at 300 kV and equipped with an EDX spectrometer. The focused ion beam (FIB) technique was used to prepare the samples for cross-sectional imaging.
The measurements SIMS (Secondary Ion Mass Spectrometry) were carried out using the CAMECA IMS6F system. A Cs+ primary beam with an energy of 5.5 keV and a constant current of 100 nA was microscope (SEM) equipped with a Gatan MonoCL 3 cathodoluminescence (CL) system and liquid helium cooled cryo-stage was used. The CL spectra of Eu-doped {ZnCdO/ZnMgO}$_{22}$ SLs before and after annealing were collected using a primary electron beam with an energy of $E_B$ = 10 keV and a current of $I_B$ = 12 nA. Transmittance spectra were studied at room temperature (RT) using a Jasco V-700 spectrophotometer in the range from 300 to 700 nm.

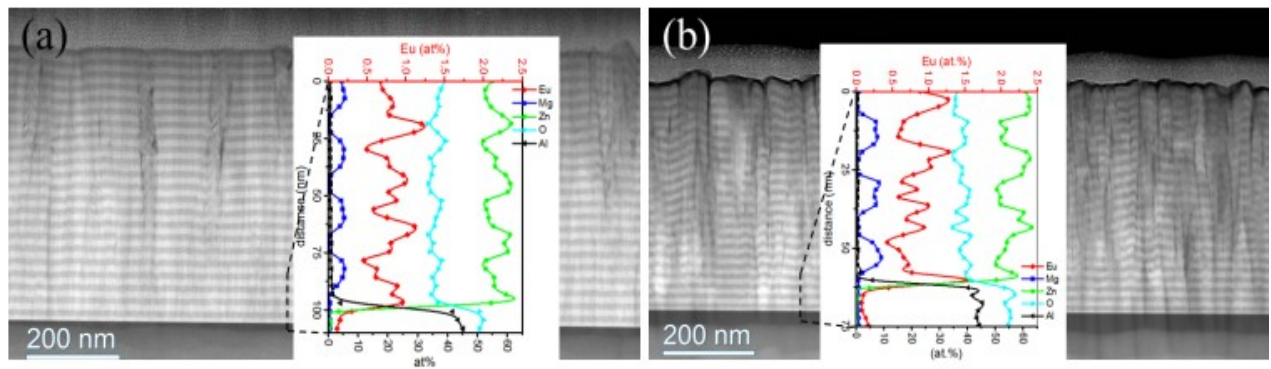

Fig. 1. The STEM/HAADF images of (a) samples A {ZnCdO:Eu$_{13nm}$/ZnMgO$_{12.5nm}$}$_{22}$ SL and (b) samples B {ZnCdO:Eu$_{10.5nm}$/ZnMgO$_{10nm}$}$_{22}$ SL together with line profiles of elements' concentration along the growth direction.

## 3. Results

### 3.1 TEM analysis

Cross-sectional high-angle annular dark-field scanning transmission electron microscopy (HAADF/STEM) images of as-grown Eu-doped {ZnCdO/ZnMgO}$_{22}$ SLs samples A and B are shown in Fig 1. The images demonstrate a well-defined periodic layered structures consisting of arrays of bright and dark stripes corresponding to ZnCdO:Eu QWs and ZnMgO barriers, respectively. The first few layers are almost perfect, but as the thickness increases, disturbances appear in the form of vertical blocks. The sample A shows a well-defined layered structure consisting of vertical blocks approximately 150 nm wide in contrast to sample B, which reveals a worse morphology and a larger number of vertical blacks about 50 nm wide. It is worth noting the difference in the surface roughness visible in Fig 1: ~14.34 nm for sample A and ~32.88 nm for sample B. The Eu and Zn concentration profiles in the graphs have common maxima, while the Mg concentration is slightly higher for sample B. The layer thicknesses are presented in Table 1.

No mismatch dislocations were found at the layer interfaces, but the existence of threading or other dislocations on at the block boundaries is not excluded.
The lack of mismatch dislocations at the interfaces between the layers suggests the existence of stresses inside these layers [Fig 2]. The Geometric Phase Analysis (GPA) method was used to examine the strain in the growth direction i.e. for (10.1) interplanar distances in areas of single blocks [in Fig. 2 marked A, B and C][27,28]. The analysis showed the existence of compressive strain for ZnMgO barriers and tensile strain for ZnCdO:Eu QWs and the difference between the strains in the barriers and QWs does not exceed approximately 2%. Due to a large number of vertical blocks, determining the strains in sample B is subject to a much greater error [Fig. 2(b)].

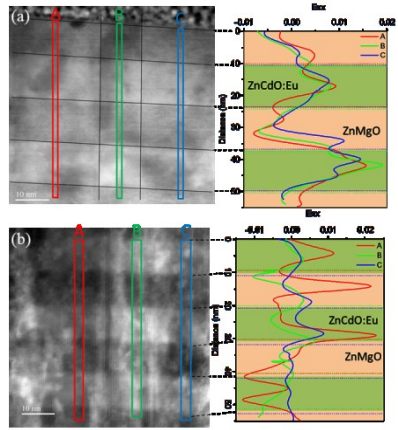

Fig. 2. STEM/HAADF images of (a) sample A and (b) sample B together with strain profiles in the growth direction. The stress profiles in the growth direction were analysed in the areas marked A (red), B (green), and C (blue) and their values are plotted on the right side of the figures.

### 3.2 XRD analysis

Fig. 3(a) presents XRD patterns of as-deposited Eu-doped {ZnCdO/ZnMgO}$_{22}$ SLs grown on $m$-plane Al$_2$O$_3$ substrate. The XRD patterns reveal that samples A and B have the hexagonal structure of ZnO (JCPDS Card 00-005-0664). For both samples the XRD signal coming from the $m$-plane sapphire substrate (JCPDS Card 00-050-0792) [marked as (*) in Fig. 3(a)] was recorded. Moreover, for sample A only the wurtzite 10.0, 20.0 and 30.0 diffracted intensities were observed. For sample B diffraction peaks from additional wurtzite orientations are indicated, namely 00.2, 11.0. Diffraction peaks associated with impurities or secondary phases (rocksalt CdO, MgO and/or Eu$_2$O$_3$) were not observed.

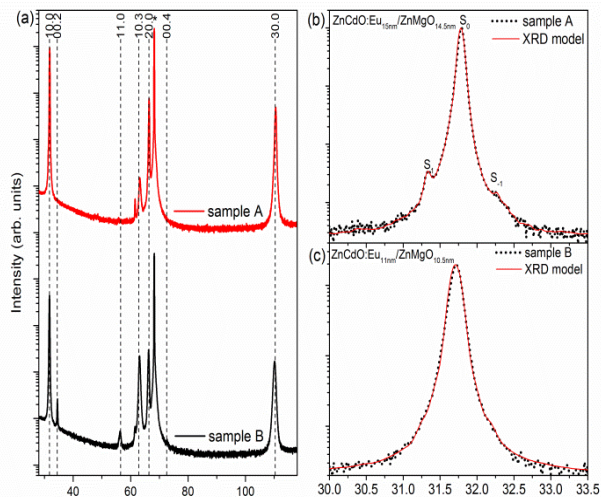

Fig. 3(a) Low resolution XRD patterns of as-grown {ZnCdO:Eu/ZnMgO}22 SLs with varying Zn content. The black vertical dotted lines correspond to the peak positions of wurtzite ZnO and asterisks (*) indicates the peak originated from the Al$_2$O$_3$ substrate; (b) High resolution XRD 2θ-ω scans of the 10.0 peaks for as-grown {ZnCdO:Eu$_{13.0nm}$/ZnMgO$_{12.5nm}$}$_{22}$ SL and (c) for as-grown {ZnCdO:Eu$_{10.5nm}$/ZnMgO$_{10nm}$}$_{22}$ SL. The XRD experimental data are shown as black lines and the simulation results as red lines.

Additionally, for both samples a diffracted signal marked as 10.3 was registered. It corresponds to the presence of a twin structure in samples A and B. Similar behavior was already observed in our previous works[29,30] and it is described in the literature for $m$-plane ZnO sample[31–33]. Using the expression[30]

$$I^*_{(hkl)} = \frac{I_{(hkl)}}{I^o_{(hkl)}}. \tag{1}$$

where $I_{(hkl)}$ is the measured relative intensity, $I^o_{(hkl)}$ is the standard signal intensity of the $(hkl)$ plane taken from the JCPDS data, the ratios of 10.0 and 10.3 peak intensities were obtained. For the sample A, the intensity $I_{10.0}/I_{10.3}$ the ratio is ~314 while for sample B is ~10. The calculated ratios indicate fewer amounts of twins for sample A compared to sample B. In sample B the crystallographic structure has greater disorder therefore the intensity ratio is more than 30 times smaller than in sample A.

For a thorough analysis of the examined structures high resolution X-ray diffraction measurements were performed. 2θ/ω scans were presented on Fig. 3(b, c). For sample A, characteristic satellite peaks ($S_1$ and $S_{-1}$) surrounding the main peak ($S_0$) were observed, confirming the good quality of the periodic superlattice structures [Fig. 3(b)]. For sample B [Fig. 3(c)] the satellite peaks coming from the SL are hidden in the zeroth order peak ($S_0$). This is related to the worse quality of this SL and the broadening of the $S_0$ peak. Theoretical XRD data were modeled using the MROX software [red lines in Fig. 3(b, c)] and they correlate perfectly with the experimental XRD curves [black lines in Fig. 3(b, c)]. Simulation of the XRD data allowed us to obtain average thicknesses of individual ZnCdO:Eu and ZnMgO sublayers. These values agree well with the thicknesses observed in the TEM images (Table 1). The absence of satellite peaks for sample B is explained by the worse crystallographic quality of the SL, occurrence of additional planes orientation and also by the design structure (composition and sublayer thickness), as confirmed by simulations (Fig. 3(c)). Also, the broadening of satellite peaks and loss of long-range order may lead to deformations in the structure, disorder at the interfaces and variations in the thickness of individual layers[34].

The accurate values of the lattice constant $a$ and $c$ for the as-deposited superlattice structures were calculated from the HRXRD maps (not shown here) using a 20.0 symmetrical reflection and a 20.-3 asymmetrical reflection. The obtained values are presented in Table 1. Description of the procedure of measurements of lattice parameters with uncertainties was proposed by Fewster[35].

The lattice constants differ slightly in comparison to the lattice constants given in the standard data for bulk ZnO ($a_0 = 3.249$ Å and $c_0 = 5.205$ Å, JCPDS 00-005-0664). For both samples we obtain smaller value of the *c*-lattice parameters and higher value of the *a*-lattice parameters. The higher difference in the lattice parameter was observed for sample B. Kumar et al.[5] reported an increase in the lattice parameter in ZnO:Eu$^{3+}$ nanophosphors, correlating with higher Eu doping concentrations. This is due to the incorporation of Eu ions into the ZnO-host matrix, which have a larger ionic radius than Zn ions[5].

These ions occupy Zn$^{2+}$ lattice sites, thus leading to the expansion of interatomic distances and consequently, to an increase in the lattice constants.. Additionally, an increase in the lattice parameters may be due to the presence of Cd$^{2+}$ ions in the ZnO wurtzite lattice[9]

### 3.2 Band gap energy

The band gap ($E_g$) of as-grown Eu-doped {ZnCdO/ZnMgO}$_{22}$ SLs was determined from the transmission spectra [Fig. 4(a)] using the Tauc's relation[36]

$$\alpha h\nu = \beta(h\nu - E_g)^n, \tag{2}$$

where α is the absorption coefficient, $h\nu$ is the photon energy, $\beta$ is a constant, and $n$ is a number characterizing the transition process. For allowed direct transitions[37], $n = 1/2$, and for allowed indirect transitions, $n = 2$. The optical band gap was estimated by extrapolating the linear regions of the Tauc's plot (($\alpha h\nu)^2 = f(h\nu)$) to zero [Fig. 4(b)]. The values of band gaps are listed in Table 1.

It is well-known from the literature that an increase in the Eu concentration in ZnO and CdMgZnO samples typically leads to a reduction in the band gap[1,14,38]. However, in superlattices, the band gap is influenced by a complex interaction of multiple factors, including sublayers composition and their thicknesses, lattice deformations and mismatches between layers and also quantum effects[29,39]. In our case, a smaller value of the band gap is observed for the superlattice with the thinnest ZnCdO:Eu and ZnMgO sublayers (sample B).

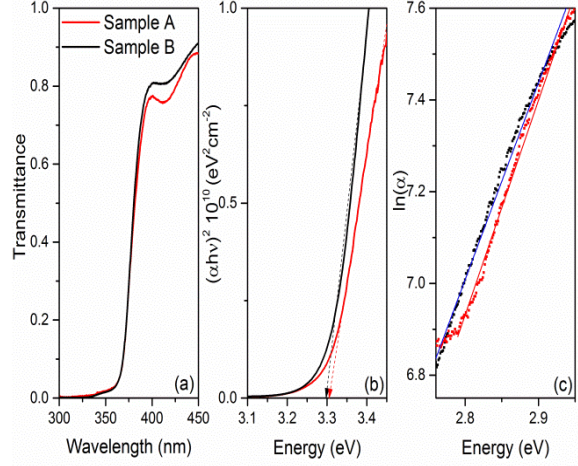

Fig. 4 (a) The normalized transmission spectra; (b) Tauc's plots of as-grown samples; (c) semi-logarithmic dependence of the absorption coefficient on photon energy for {ZnCdO:Eu/ZnMgO}$_{22}$ and {ZnCdO:Eu/ZnMgO}$_{22}$ SLs.

The Urbach energy ($E_u$) related to the tail of states localized within the band gap, can be found using the following dependence[40,41],

$$\alpha = \alpha_0 exp\left(\frac{h\nu}{E_u}\right), \qquad (3)$$

where $\alpha_0$ is a constant. The value of $E_u$ was estimated from the inverse dependence of the slope $ln(\alpha)$ on $(h\nu)$ as shown in Fig. 4(c)). The values of 214 meV and 233 meV were obtained for sample A and B, respectively (Table 1). A higher Urbach energy value is observed for sample B, which is consistent with TEM and XRD measurements that indicate a worse periodic structure in this sample compared to sample A.

### 3.3 Elemental composition

SIMS depth profiles of Eu-doped {ZnCdO/ZnMgO}$_{22}$ SLs before and after annealing at different temperatures are shown in Fig. 5. The depth profiles clearly reveal the presence of alternating layers of ZnCdO:Eu and ZnMgO in the superlattice structures. As expected, each Cd and Eu peak is accompanied by a minimum of the Mg peak at the same depth position. The oscillations of the Eu, Cd and Mg signals, are repeated 22 times in agreement with the designed structures. The periodicity of elemental signals is more pronounced in the SIMS depth profiles of as-grown {ZnCdO:Eu$_{13nm}$/ZnMgO$_{12.5nm}$}$_{22}$ SL (sample A). This observation is in good agreement with the TEM measurements of the as-deposited sample A [see Fig. 1(a)]. The lack of periodicity in the SIMS signal in the {ZnCdO:Eu$_{10.5nm}$/ZnMgO$_{10nm}$}$_{22}$ SL (sample B) is due to the lower depth resolution of SIMS measurement caused by the increased surface roughness seen in TEM measurement [Fig. 1(b)]. Undulation of the Zn and O profiles is a measurement artifact associated with the matrix effect that occurs when measuring a stepwise changing matrix element content[42]. Previous studies have demonstrated the clear identification of individual CdO and ZnO layers in SIMS depth profiles in as-deposited {CdO/ZnO}$_m$ SLs on $m$-plane Al$_2$O$_3$ using the MBE method[43].

After annealing the SL structures at various temperatures in an O$_2$ atmosphere for 1 minute, the oscillations intensity in the SIMS depth profile for Mg, Cd and Eu elements decrease with rising RTP temperatures. Less pronounced oscillations of SIMS profiles indicate the mutual diffusion of all elements. It should be noted that diffusion is more active in the ZnCdO:Eu sublayers. At 900°C, the periodic structure of superlattices almost disappears, and the corresponding SIMS depth profiles showed the flattened lines for the Cd and Eu ion signals. Weak oscillations remain visible for the Mg depth profile.

In Zn(Cd,MgS)Se/ZnSe SL structures, the diffusion process of Mg and Cd elements was studied, and it was found that Mg ions were more stable than Cd ions[38,44]. It was noted that the Mg diffusion coefficient is orders of magnitude smaller than the diffusion coefficient of Cd under comparable conditions[38]. Cd segregation at the layer-substrate interface was observed in {CdO/ZnO}$_m$ SLs after annealing at 900°C for 5 minutes in an O$_2$ atmosphere[43]. Meanwhile, the homogeneity of Cd distribution in annealed periodical structures depended on the thickness of the CdO and ZnO layers.

Table 1. Thickness of sublayers, lattices parameters, Urbach and band gap energy of as-grown {ZnCdO:Eu/ZnMgO}$_{22}$ SLs

| Sample | TEM | | XRD | | | | | |
|---|---|---|---|---|---|---|---|---|
| | $h_{ZnCdO:Eu}$ ±1 (nm) | $h_{ZnMgO}$ ±1 (nm) | $h_{ZnCdO:Eu}$ ±1 (nm) | $h_{ZnMgO}$ ±1 (nm) | $a$ ±0.003 (Å) | $c$ ±0.003 (Å) | $E_g$ ±0.005 (eV) | $E_U$ ±5 (meV) |
| A | 13 | 12.5 | 10.5 | 10.5 | 3.252 | 5.203 | 3.318 | 214 |
| B | 10.5 | 10 | 10 | 10 | 3.254 | 5.171 | 3.306 | 233 |

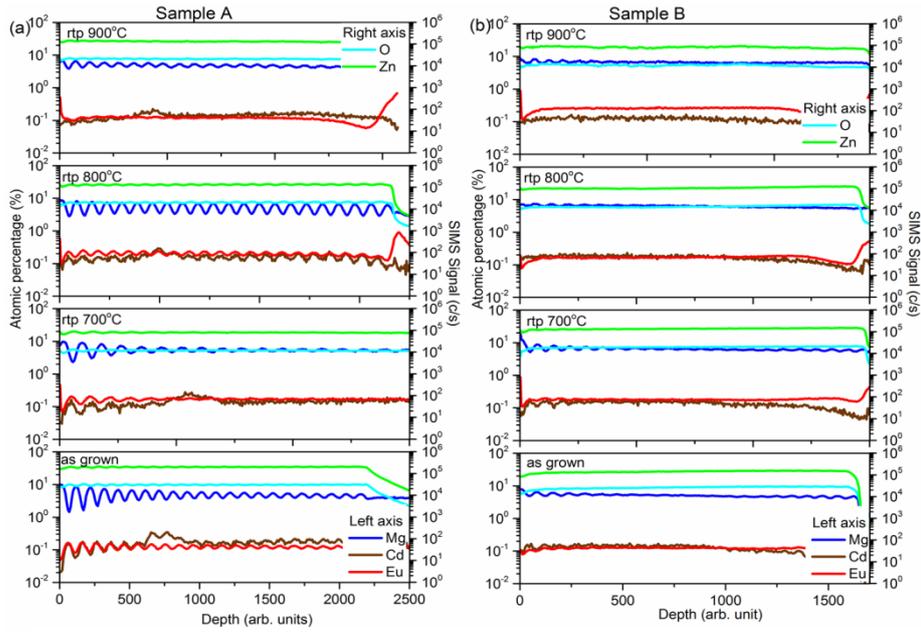

Fig. 5. SIMS depth profiles of as-grown and annealed at different temperatures SLs: (a) {ZnCdO:Eu$_{13nm}$/ZnMgO$_{12.5nm}$}$_{22}$ SL and (b) {ZnCdO:Eu$_{10.5nm}$/ZnMgO$_{10nm}$}$_{22}$.

### 3.4 Optical properties

Cathodoluminescence (CL) spectra were obtained for both as-deposited and annealed {ZnCdO:Eu/ZnMgO}$_{22}$ SLs at low (LT ~5 K) and at room (RT) temperatures. The normalized CL spectra presented in Fig. 6 are similar to each other. All CL spectra demonstrate a dominant near-band-edge (NBE) emission peak at 377±4 nm associated with radiative recombination of free excitons[45]. In addition to the NBE emission, a broad visible luminescence covering the range of 430-660 nm is observed. Typically, broadband emission in the visible region is associated with deep level structural defects such as oxygen vacancies ($V_O$), oxygen interstitials ($O_i$), oxygen antisites ($O_{Zn}$), zinc vacancies ($V_{Zn}$), zinc interstitials ($Zn_i$) or zinc antisites ($Zn_O$)[5,46]. In the low-energy region of the spectra, well resolved luminescence lines, due to the $^5D_1 \rightarrow ^7F_J$ (J = 0-4) Eu$^{3+}$ intra-4f$^6$ shell transitions are visible [3,14].

The high-energy peak at ~344±1 nm corresponds to the emission originating from the ZnMgO barrier layer[33]. This peak is more visible at low temperatures than at room temperature (Fig. 6). For the {ZnCdO:Eu$_{13nm}$/ZnMgO$_{12.5nm}$}$_{22}$ SL (sample A), the peak disappears after RTP above 700°C. In contrast, for the {ZnCdO:Eu$_{10.5nm}$/ZnMgO$_{10nm}$}$_{22}$ SL (sample B), the peak intensity decreases, but it remains visible at low temperatures even after annealing at 900°C. For sample B, the NBE peak shifted to higher energy in comparison with the peak position for sample A. This shift, as well as the emission from the ZnMgO barrier layer after annealing at 900°C, may be associated with a higher Mg content in the structure B[8]. The full width at half maximum (FWHM) of the dominant NBE peak increases with rising annealing temperature and a stronger effect is observed for sample A compared to sample B. The lower FWHM value observed for as-deposited superlattices may indicate a smaller number of structural defects and better optical qualities of these samples[47].

In the CL spectra in Fig. 6(a) a broad emission band from 430 to 660 nm, originating from structural defects in the superlattices is observed. The green band at ~545 nm is probably the result of electron recombination in single ionized oxygen vacancies with photo excited holes in the valence band[45]. For example, Ton-That et al.[48] showed that for Zn- and O-rich ZnO particles, the green emission bands in the luminescence spectra are located at ~492 nm and ~539 nm. Pan et al.[13] attributed the green luminescence at ~515 nm in ZnO:Eu nanowires to Eu impurities. The intensity of the deep level emission (DLE) of {ZnCdO:Eu/ZnMgO}$_{22}$ samples changed with the annealing temperature. For example, Lee et al. observed for ZnO films grown on silicon wafers using an RF magnetron sputtering after RTP conducted in an O$_2$ atmosphere at a pressure of 1 Torr at temperatures from 500 to 1000°C for 3 minutes, an increase in the intensity of peaks related to $O_i$ and $O_{Zn}$ defects above 900°C, while the peak associated with $V_O$ practically did not change its intensity over the entire annealing temperature range[49]

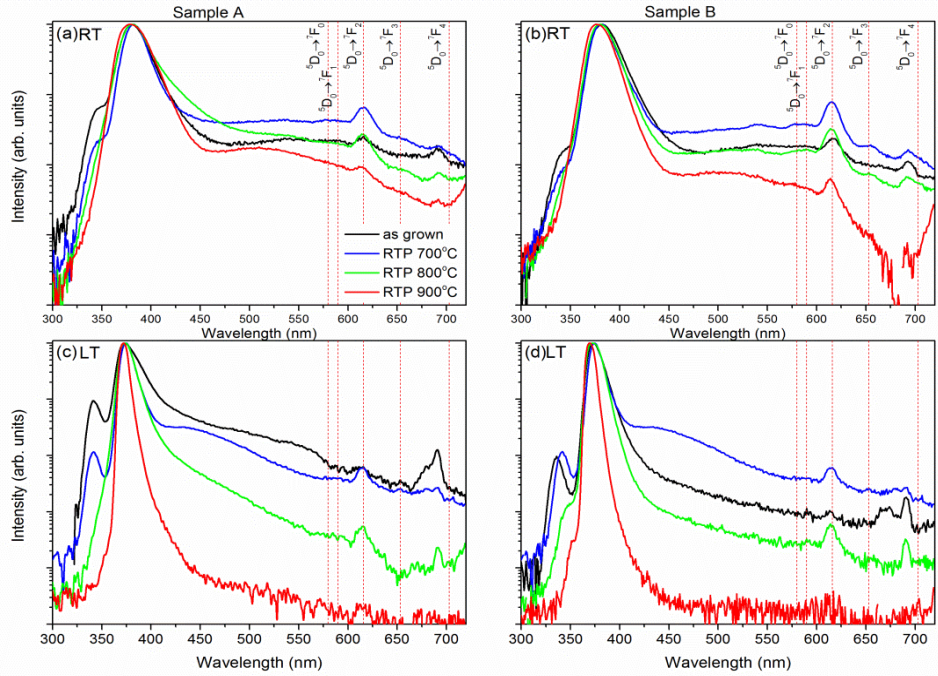

Fig. 6. Semi-logarithmic plot of the normalized CL spectra of as-grown and annealed (a) {ZnCdO:Eu$_{13nm}$/ZnMgO$_{12.5nm}$}$_{22}$ SL at RT; (b) {ZnCdO:Eu$_{10.5nm}$/ZnMgO$_{10nm}$}$_{22}$ SL at RT; (c) {ZnCdO:Eu$_{13nm}$/ZnMgO$_{12.5nm}$}$_{22}$ SL at LT; and (d) {ZnCdO:Eu$_{10.5nm}$/ZnMgO$_{10nm}$}$_{22}$ SL.

Fig. 6 also shows the typical peak positions of the Eu$^{3+}$ emission due to the $^5D_0 \rightarrow {}^7F_{0,1,2,3,4}$ intra-4f-shell transitions (580, 590, 616, 653, and 703 nm)[14]. The red emission line at ~616 nm due to the $^5D_0 \rightarrow {}^7F_2$ stimulated electric dipole transition mechanism dominates in all samples[5,13]. The intensity of $^5D_0 \rightarrow {}^7F_2$ transition increases as the symmetry of the Eu$^{3+}$ centers decreases[5]. The appearance of characteristic intra-4f shell transitions of RE is associated with host-guest energy transfer (ET), probably involving the ZnO defect states. Layek et al.[46] attributed the increase in the intensity of visible luminescence in RE$^{3+}$ ions doped ZnO nanorods to a large number of defects resulting from the incorporation of RE$^{3+}$ ions into the ZnO lattice. As a consequence, energy transfer occurs from the defect states of ZnO to the RE$^{3+}$ guest ions.

Fig. 6 shows that annealing at 700°C leads to an increase in the intensity of the 616 nm line. For sample A, an increase in the annealing temperature leads to a decrease in the red emission intensity. For sample B, the emission intensity originating from Eu$^{3+}$ decreased after annealing at 800°C. Suzuki et al.[12] also observed a gradual decrease in the red emission intensity for Eu-doped ZnO nanocrystalline material as the annealing temperature increased from 533 to 633K. They suggested that the charge-transfer (CT) band shifts to higher energies regions with increasing annealing temperature, and may merge with the conduction band. This shift was explained by the increase in the band gap of ZnO nanoparticles and covalency between Eu and O atoms. The efficiency of energy transfer from oxygen atoms in the host crystals to the 4f-orbital states of Eu ions via CT is probably increased due to the fact that the CT band partially overlaps the band gap. Comparing the CL spectra, we might suggest that the presence of defects increases the efficiency of energy transfer from the host lattice to the luminescence centers.

The black points in Fig. 7 shows the ratio of NBE to DLE emission intensities as a function of the annealing temperature for both samples at 300K. It should be noted, that the $I_{NBE}/I_{DLE}$ ratio reveals a similar trend in both structures. The red points in Fig. 7 show the change in the integrated $^5D_0 \rightarrow {}^7F_2$ peak area at RT with the change in annealing temperature. The choice of $I_{NBE}/I_{DLE}$ ratio and the peak area can be justified. The highest intensity of the red line is observed for the lowest $I_{NBE}/I_{DLE}$ ratio, which may indicate active participation in energy transfer from the host material to Eu ions through defect states[15,46,50]. Based on the analysis of CL spectra, the electronic energy level diagram for the {ZnCdO:Eu/ZnMgO}$_{22}$ SLs is illustrated in Fig. 8.

The International Commission on Illumination (CIE) color coordinates $(x, y)$ for both superlattice structures at RT are listed in Table 2 and plotted on the CIE chromaticity diagram, as shown in Fig. 9. The CIE parameters for 2 samples after annealing at 700°C are located in the orange-red region of the diagram and confirm the conclusions made after analyzing the CL spectra (see Fig. 6).

Analysis of the CL spectra (Fig. 6 and Fig. 9) for both structures suggests that red emission is more effective for superlattices with thinner sublayers, (sample B). This effect could be attributed to quantum effects in the {ZnCdO:Eu$_{10.5nm}$/ZnMgO$_{10nm}$}$_{22}$ SL or to the higher Mg concentration in the ZnMgO barriers, which improves the Eu$^{3+}$ emission. Additionally, sample B has comparatively poorer crystallographic quality as shown by TEM and XRD measurements as well as a higher Urbach parameter compared to sample A. Consequently, a greater number of structural defects may be involved in energy transport between the host-matrix and Eu$^{3+}$ ions in the {ZnCdO:Eu$_{10.5nm}$/ZnMgO$_{10nm}$}$_{22}$ SL. The detailed mechanisms of europium emission in superlattices, in particular ZnCdO/ZnMgO structures, require further investigation.

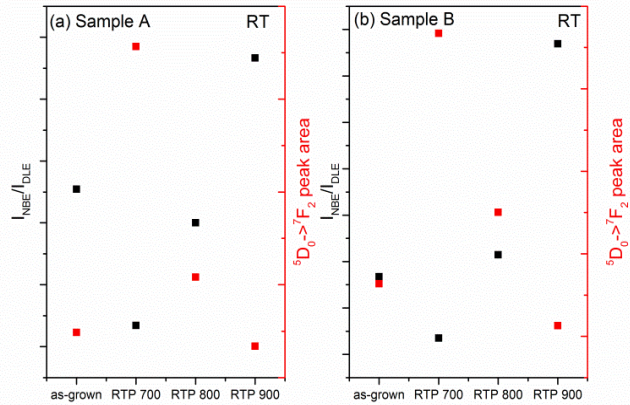

Fig. 7. $I_{NBE}/I_{DLE}$ ratio and integrated $^5D_0 \rightarrow {}^7F_2$ peak area as a function of annealing temperature.

Table 2. CIE coordinates of Eu-doped {ZnCdO/ZnMgO}$_{22}$ SLs before and after annealing in an O$_2$ for 1 minute at room temperature.

| Room temperature | | | |
|---|---|---|---|
| Sample | | $x$ | $y$ |
| Sample A | As-grown | 0.3 | 0.275 |
| | RTP 700°C | 0.321 | 0.307 |
| | RTP 800°C | 0.242 | 0.195 |
| | RTP 900°C | 0.246 | 0.223 |
| Sample B | As-grown | 0.293 | 0.257 |
| | RTP 700°C | 0.34 | 0.313 |
| | RTP 800°C | 0.308 | 0.271 |
| | RTP 900°C | 0.267 | 0.246 |

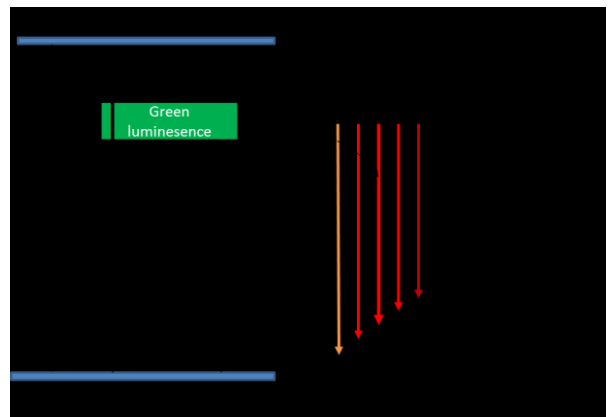

Fig. 8. Schematic band diagram of in situ Eu-doped {ZnCdO/ZnMgO}$_{22}$ SLs.

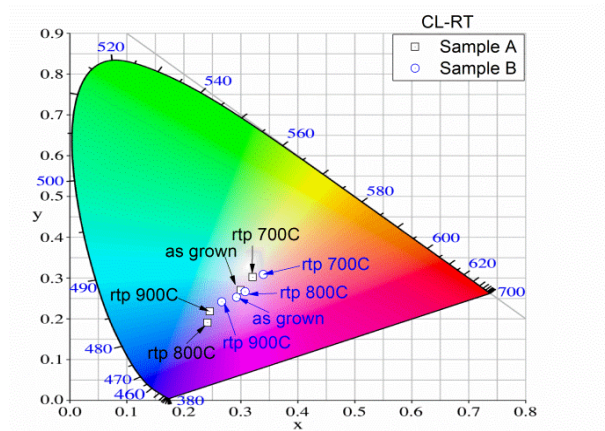

Fig. 9. The CIE chromaticity diagram for in situ Eu-doped {ZnCdO/ZnMgO}$_{22}$ SLs

## Conclusions

In situ Eu-doped {ZnCdO/ZnMgO}$_{22}$ SLs with varying sublayer thicknesses were successfully deposited by plasma-assisted molecular beam epitaxy on *m*-plane Al$_2$O$_3$ substrates. The structural and optical properties of these samples were characterized using XRD, TEM, UV–Vis spectroscopy, and CL techniques. We observed that the structure with wider ZnCdO:Eu and ZnMgO layers has better crystallographic quality
compared to the superlattice with thinner sublayers. This result was confirmed by the analysis of TEM and XRD measurements. A deformation of ~1% in the growth direction (10.1) was found for both structures. SIMS depth profiles for as grown samples demonstrate fluctuations in the Eu, Cd, Zn and Mg signals reflecting the periodic nature of the structures. In the as-grown SLs, emission in the UV region dominates, and the emission line at 616 nm corresponding to the transition $^5D_0 \rightarrow {^7F_2}$ is also observed. To optimize the emission efficiency of Eu$^{3+}$ ions, the samples were annealed at several temperatures under oxygen flow. The amplitude of SIMS signals is reduced with increasing temperature due to interdiffusion of elements. In the CL spectra, we observe an increase in the Eu emission after annealing at 700°C, while annealing at higher temperatures leads to the opposite effect. Consequently, more intense red emission was observed in the superlattice with thinner sublayers than with wider ones.

## Conflicts of Interest

The authors declare no conflict of interest

## Acknowledgements

## Acknowledgements

This work was supported in part by the Polish National Science Center, Grants No. 2021/41/B/ST5/00216 and 2019/35/B/ST8/01937,.

The author S. Magalhaes was suporrted by Portuguese Foundation for Science and Technology (FCT) in the framework of the Plurianual Strategic Funding UIDB/50010/2020, UIDP/50010/2020, LA/P/0061/2020.

The author M. Schot was suporrted by "MagTop" project (FENG.02.01-IP.05-0028/23) carried out within the "International Research Agendas" programme of the Foundation for Polish Science co-financed by the European Union under the European Funds for Smart Economy 2021-2027 (FENG).

## Notes and references

[1]  O. Kalu, I. Ahemen, H. E. Ponce, J. Alberto, D. Moller, *J. Phys. D. Appl. Phys.* **2021**, *54*, 345108.
[2]  N. us Saqib, R. Adnan, I. Shah, *Environ. Sci. Pollut. Res.* **2016**, *23*, 15941.
[3]  A. Kumawat, K. P. Misra, S. Chattopadhyay, *Mater. Technol.* **2022**, *37*, 1595.
[4]  S. W. Choi, S. Emura, S. Kimura, M. S. Kim, Y. K. Zhou, N. Teraguchi, A. Suzuki, A. Yanase, H. Asahi, *J. Alloys Compd.* **2006**, *408–412*, 717.
[5]  V. Kumar, V. Kumar, S. Som, M. M. Duvenhage, O. M. Ntwaeaborwa, H. C. Swart, *Appl. Surf. Sci.* **2014**, *308*, 419.
[6]  J. A. Mathew, V. Tsiumra, J. M. Sajkowski, A. Wierzbicka, R. Jakiela, Y. Zhydachevskyy, E. Przezdziecka, M. Stachowicz, A. Kozanecki, *J. Lumin.* **2022**, *251*, 119167.
[7]  S. Singh, P. Thiyagarajan, K. Mohan Kant, D. Anita, S. Thirupathiah, N. Rama, B. Tiwari, M. Kottaisamy, M. S. Ramachandra Rao, *J. Phys. D. Appl. Phys.* **2007**, *40*, 6312.
[8]  A. El-Shaer, A. Bakin, M. Al-Suleiman, S. Ivanov, A. Che Mofor, A. Waag, *Superlattices Microstruct.* **2007**, *42*, 129.
[9]  O. Kalu, J. A. Duarte Moller, A. Reyes Rojas, *Phys. Lett. Sect. A Gen. At. Solid State Phys.* **2019**, *383*, 1037.
[10] L. F. Koao, B. F. Dejene, H. C. Swart, S. V. Motloung, T. E. Motaung, *Opt. Mater. (Amst).* **2016**, *60*, 294.
[11] T. Ghrib, A. L. Al-Otaibi, I. Massoudi, A. M. Alsagry, A. S. Aljaber, E. A. Alhussain, W. S. Alrubian, S. Brini, M. A. Gondal, K. A. Elsayed, T. S. Kayed, *Arab J. Basic Appl. Sci.* **2022**, *29*, 138.
[12] K. Suzuki, K. Murayama, N. Tanaka, *Appl. Phys. Lett.* **2015**, *107*, 031902.
[13] C. J. Pan, C. W. Chen, J. Y. Chen, P. J. Huang, G. C. Chi, C. Y. Chang, F. Ren, S. J. Pearton, *Appl. Surf. Sci.* **2009**, *256*, 187.
[14] E. Hasabeldaim, O. M. Ntwaeaborwa, R. E. Kroon, H. C. Swart, *J. Mol. Struct.* **2019**, *1192*, 105.
[15] L. Yang, Z. Jiang, J. Dong, A. Pan, X. Zhuang, *Mater. Lett.* **2014**, *129*, 65.
[16] A. Kozanecki, J. M. Sajkowski, J. A. Mathew, Y. Zhydachevskyy, E. Alves, M. Stachowicz, *Appl. Phys. Lett.* **2021**, *119*, 112101.
[17] J. L. Frieiro, C. Guillaume, J. López-Vidrier, O. Blázquez, S. González-Torres, C. Labbé, S. Hernández, X. Portier, B. Garrido, *Nanotechnology* **2020**, *31*, 465207.
[18] H. J. Lozykowski, W. M. Jadwisienczak, J. Han, I. G. Brown, *Appl. Phys. Lett.* **2000**, *77*, 767.
[19] S. Magalhães, M. Peres, V. Fellmann, B. Daudin, A. J. Neves, E. Alves, T. Monteiro, K. Lorenz, *J. Appl. Phys.* **2010**, *108*, 084306.
[20] N. Ben Sedrine, J. Rodrigues, D. N. Faye, A. J. Neves, E. Alves, M. Bockowski, V. Hoffmann, M. Weyers, K. Lorenz, M. R. Correia, T. Monteiro, *ACS Appl. Nano Mater.* **2018**, *1*, 3845.
[21] M. Brahlek, A. Sen Gupta, J. Lapano, J. Roth, H. T. Zhang, L. Zhang, R. Haislmaier, R. Engel-Herbert, *Adv. Funct. Mater.* **2018**, *28*, 1702772.
[22] K. He, *Chinese Phys. B* **2022**, *31*, 126804.
[23] M. H. Asghar, F. Placido, S. Naseem, *Eur. Phys. JournalApplied Phys.* **2006**, *184*, 177.
[24] M. A. Ebdah, M. E. Kordesch, A. Anders, W. M. Jadwisienczak, *Mater. Res. Soc. Symp. Proc.* **2010**, *1202*, 269.
[25] S. Magalhães, J. S. Cabaço, J. P. Araújo, E. Alves, *CrystEngComm* **2021**, *23*, 3308.
[26] S. Magalhães, C. Cachim, P. D. Correia, F. Oliveira, F. Cerqueira, J. M. Sajkowski, M. Stachowicz, *CrystEngComm* **2023**, *25*, 4133.
[27] M. Takeda, J. Suzuki, *JOSA A* **1996**, *13*, 1495.
[28] H. Du, **2018**.
[29] E. Przeździecka, A. Wierzbicka, A. Lysak, P. Dłużewski, A. Adhikari, P. Sybilski, K. Morawiec, A. Kozanecki, *Cryst. Growth Des.* **2022**, *22*, 1110.
[30] A. Lysak, E. Przeździecka, A. Wierzbicka, P. Dłużewski, J. Sajkowski, K. Morawiec, A. Kozanecki, *Cryst. Growth Des.* **2023**, *23*, 134.
[31] J.-H. Kim, S. K. Han, S. I. Hong, S.-K. Hong, J. W. Lee, J. Y. Lee, J.-H. Song, J. S. Park, T. Yao, *J. Vac. Sci. Technol. B Microelectron. Nanom. Struct.* **2009**, *27*, 1625.
[32] H. Tanoue, M. Takenouchi, T. Yamashita, S. Wada, Z. Yatabe, S. Nagaoka, Y. Naka, Y. Nakamura, *Phys. Status Solidi Appl. Mater. Sci.* **2017**, *214*, 1600603.
[33] C. C. Kuo, B. H. Lin, S. Yang, W. R. Liu, W. F. Hsieh, C. H. Hsu, *Appl. Phys. Lett.* **2012**, *101*, 011901.
[34] H. V. Stanchu, A. V. Kuchuk, P. M. Lytvyn, Y. I. Mazur, Y. Maidaniuk, M. Benamara, S. Li, S. Kryvyi, V. P. Kladko, A. E. Belyaev, Z. M. Wang, G. J. Salamo, *Mater. Des.* **2018**, *157*, 141.
[35] P. F. Fewster, *J. Mater. Sci. Mater. Electron.* **1999**, *10*, 175.
[36] A. V. Tauc, J., Radu Grigorovici, *Phys. status solidi* **1966**, *15*, 627.
[37] A. Bedia, F. Z. Bedia, M. Aillerie, N. Maloufi, B. Benyoucef, *Energy Procedia* **2015**, *74*, 529.
[38] M. Straßburg, M. Kuttler, O. Stier, U. W. Pohl, D. Bimberg, M. Behringer, D. Hommel, *J. Cryst. Growth* **1998**, *184–185*, 465.
[39] P. Kasap, S., Capper, *Springer Handbook of Electronic and Photonic Materials*, Springer, **2017**.
[40] Regular, K. Boubaker, **2011**, *126*, 1.
[41] P. Bindu, S. Thomasa, *Acta Phys. Pol. A* **2017**, *131*, 1474.

[42]   Z. Khosravizadeh, P. Dziawa, S. Dad, R. Jakiela, *Thin Solid Films* **2023**, *781*, 139974.
[43]   A. Lysak, E. Przeździecka, R. Jakiela, A. Reszka, B. Witkowski, Z. Khosravizadeh, A. Adhikari, J. M. Sajkowski, A. Kozanecki, *Mater. Sci. Semicond. Process.* **2022**, *142*, 1.
[44]   M. Kuttler, M. Strassburg, V. Türck, R. Heitz, U. W. Pohl, D. Bimberg, E. Kurtz, G. Landwehr, D. Hommel, *Appl. Phys. Lett.* **1996**, *69*, 2647.
[45]   R. Raji, K. G. Gopchandran, *J. Sci. Adv. Mater. Devices* **2017**, *2*, 51.
[46]   A. Layek, S. Banerjee, B. Manna, A. Chowdhury, *RSC Adv.* **2016**, *6*, 35892.
[47]   M. A. Boukadhaba, A. Fouzri, V. Sallet, S. S. Hassani, G. Amiri, A. Lusson, M. Oumezzine, *Superlattices Microstruct.* **2015**, *85*, 820.
[48]   C. Ton-That, L. Weston, M. R. Phillips, *Phys. Rev. B - Condens. Matter Mater. Phys.* **2012**, *86*, 1.
[49]   D. K. Lee, S. Kim, M. C. Kim, S. H. Eom, H. T. Oh, S. H. Choi, *J. Korean Phys. Soc.* **2007**, *51*, 1378.
[50]   S. M. Ahmed, P. Szymanski, L. M. El-Nadi, M. A. El-Sayed, *ACS Appl. Mater. Interfaces* **2014**, *6*, 1765.